\documentclass[a4paper]{article}

\usepackage{INTERSPEECH2019}

\usepackage{cite}
\usepackage{url}
\usepackage{enumitem}
\setlist[itemize]{leftmargin=*}

\title{Towards Generalized Speech Enhancement\\ with Generative Adversarial Networks}
\name{Santiago Pascual$^1$, Joan Serr\`a$^2$, Antonio Bonafonte$^{1\ast}$\thanks{$^\ast$A.~Bonafonte is currently at Amazon Research, Cambridge, UK.}}
\address{
  $^1$Universitat Polit\`ecnica de Catalunya\\
  $^2$Telef\'onica Research}
\email{santi.pascual@upc.edu}

\usepackage{xcolor}
\usepackage{amsmath,amssymb}
\DeclareMathOperator{\E}{\mathbb{E}}
\usepackage{mathrsfs}
\newcommand{\ve}[1]{\textbf{#1}}		

\begin{document}

\maketitle
\begin{abstract}
The speech enhancement task usually consists of removing additive noise or reverberation that partially mask spoken utterances, affecting their intelligibility. However, little attention is drawn to other, perhaps more aggressive signal distortions like clipping, chunk elimination, or frequency-band removal. Such distortions can have a large impact not only on intelligibility, but also on naturalness or even speaker identity, and require of careful signal reconstruction. In this work, we give full consideration to this generalized speech enhancement task, and show it can be tackled with a time-domain generative adversarial network (GAN). In particular, we extend a previous GAN-based speech enhancement system to deal with mixtures of four types of aggressive distortions. 
Firstly, we propose the addition of an adversarial acoustic regression loss that promotes a richer feature extraction at the discriminator. Secondly, we also make use of a two-step adversarial training schedule, acting as a warm up-and-fine-tune sequence.
Both objective and subjective evaluations show that these two additions bring improved speech reconstructions that better match the original speaker identity and naturalness.
\end{abstract}

\section{Introduction}
\label{sec:intro}
Speech enhancement comprises the improvement of intelligibility and quality of speech contaminated by noise, typically in an additive form~\cite{Loizou2013Book}. It is thus broadly applicable to scenarios that need to palliate masking artifacts over the speech, like communication systems, hearing aids, or cochlear implants, where enhancing the signal prior to amplification significantly reduces discomfort and increases intelligibility~\cite{yang2005}. Deep learning has been broadly applied to this field, either in the form of regression of waveform samples or clean spectral components, as well as prediction of masking elements over spectral bins~\cite{lu2013, xu2015regression, williamson2017time, Weninger15LVASS, erdogan2015phase, rethage2018wavenet, park2016fully}.
A recent trend in speech enhancement is also the use of generative adversarial networks (GANs)~\cite{goodfellow2014generative}, with its first application being the speech enhancement GAN (SEGAN)~\cite{pascual2017segan}. Other GAN variations have been used to avoid using aligned clean/noisy corpora~\cite{higuchi2017adversarial} or using other adversarial losses and/or domains~\cite{qin2018improved, donahue2017exploring}. SEGAN has also been applied to speech regeneration as a whispered-to-voiced alaryngeal speech conversion~\cite{pascual2018whispered}, thus extending the previous denoising approach.

Athough GANs were designed as an unsupervised learning strategy, they are proven to be more stable and effective when coupled with labels that are either injected as input conditionings~\cite{radford2015unsupervised, brock2018large} or that help classify some traits about the data being generated~\cite{odena2017conditional, lucic2019high}. Also the successful speech synthesis model parallel WaveNet~\cite{oord2017parallel} proved that a multi-task aggregation on top of the speech generative model is beneficial to improve speaker identity, prosodic, and content traits.

In this work, we first extend SEGAN towards a more generalized speech enhancement such that we recover cleaner speech out of severely distorted utterances. This generalization towards different speech distortions is interesting and directly applicable to modern communication technologies, where connections are interrupted (i.e. losing voice packets), and voice processing pipelines can distort our signals through amplifiers, compression of transmission encoders, etc. In this work we emulate these effects by introducing four applied distortions in this work, with different severity levels per distortion. Then, we propose the use of SEGAN to palliate them with three different schemes. First, a plain adversarially trained SEGAN version serves us as a baseline for our recovery experiments. Then, we propose a new regression component on top of the discriminator that serves as an additional self-supervised task, boosting the generated recovered speech quality. This new task is specially effective when we use the second modification, a two-stage adversarial training schedule as a warm up and fine-tunning sequence. Objective and subjective results show the effectiveness of both proposed mechanisms when they are jointly applied. 

\section{Speech enhancement GAN Review}

GANs define a deep generative framework, involving two neural networks and an adversarial training mechanism~\cite{goodfellow2014generative}. The two networks are the generator ($G$) and the discriminator ($D$). Both $G$ and $D$ are trained with opposite objectives, hence the adversarial nature of the framework. On the one hand, $D$ has to increasingly get better at classifying \texttt{real} or \texttt{fake} features. On the other hand, $G$ has to make up samples realistic enough to make $D$ fail in its prediction. This process is formulated as a minimax game between $G$ and $D$, and we can use different loss functions at the output of $D$ to express it. For instance a least squares loss makes it an LSGAN~\cite{Mao16ARXIV}. 
It is often interesting to add a conditioning factor to be used for a specific task, where we have some additional information $\tilde{\ve{x}}$ injected into $G$ and $D$ to perform the mapping and classification (see~\cite{Isola16ARXIV} and references therein). The formulation of $G$ and $D$ losses for a conditional LSGAN with these components is expressed as
\begin{equation}
\begin{split}
  \underset{G}\min~V(G) & = \E_{\ve{z}\sim p_{\ve{z}}(\ve{z}),\tilde{\ve{x}}\sim p_{\text{data}}(\tilde{\ve{x}})}[(D(G(\ve{z},\tilde{\ve{x}}),\tilde{\ve{x}}) - c)^{2}] ,\\
  \underset{D}\max~V(D) & = \frac{1}{3}\E_{\ve{x},\tilde{\ve{x}}\sim p_{\text{data}}(\ve{x}, \tilde{\ve{x}})}[(D(\ve{x},\tilde{\ve{x}}) - b)^{2}] \\
    & + \frac{1}{3}\E_{\ve{z}\sim p_{\ve{z}}(\ve{z}),\tilde{\ve{x}}\sim p_{\text{data}}(\tilde{\ve{x}})}[(D(G(\ve{z},\tilde{\ve{x}}),\tilde{\ve{x}}) - a)^{2}]\\ 
    & + \frac{1}{3}\E_{\ve{x},\ve{x}^{r}\sim p_{\text{data}}(\ve{x})}[(D(\ve{x},\tilde{\ve{x}}^r) - a)^{2}] ,
\end{split}
\label{eq:lsgan}
\end{equation}
where $a=-1$, $b=1$, and $c=0$ fulfill the condition to minimize Pearson $\chi^{2}$ divergence~\cite{Mao16ARXIV}. Also, $\tilde{\ve{x}}^r$ are conditioning samples unaligned with $\ve{x}$, thus creating a \texttt{fake} signal that captures that any misalignment between $\ve{x}$ and $\tilde{\ve{x}}$ is a mistake~\cite{pascual2018whispered}.

In the speech enhancement setting, we have an input noisy signal $\tilde{\ve{x}}$ that has to be cleaned to obtain the enhanced signal $\hat{\ve{x}}$. SEGAN decimates and expands the input signal feature-wise, through a number of strided convolutional layers with multiparametric rectified linear activations (PReLUs)~\cite{he2015delving}. Decimation is applied until we obtain a compact representation of a few time samples $\ve{c}$. This result is concatenated with the generative noise component $\ve{z} \sim \mathcal{N}(0, I)$, which adds the generative stochastic behavior to the generator predictions $\hat{\ve{x}}$. Decimation discourages learning a trivial identity function in the reconstruction $\tilde{\ve{x}}$, and also accelerates the convolution operations when we go deeper in the decimation structure of the encoder (shorter sequence lengths involve faster processing). It also reduces the memory footprint by using shorter feature maps. The encoding process is then reversed in the decoding stage by means of transposed convolutions, followed again by PReLUs until last layer which is $\tanh$ to bound the output between $-$1 and 1, stabilizing the adversarial process~\cite{radford2015unsupervised}. Additionally, skip connections linking the encoder and the decoder of $G$ boosted stability and generated quality.
The enhanced output signal is thus defined at the output of the decoder as $\hat{\ve{x}} = G(\ve{z},\tilde{\ve{x}})$.


Successive SEGAN works have shown its applicability to new languages with little adaptation data~\cite{pascual2018language} and its capacity to reconstruct damaged speech in the context of whispered-to-voiced alaryngeal speech conversion, also named dewhispering~\cite{pascual2018whispered}. For reconstruction, however, some details of the original architecture were modified, which led to the whisper SEGAN (WSEGAN). The most critical ones were the addition of learnable skip connections and the removal of the $L_1$ regularization term at waveform level. This eliminated a strict temporal alignment between $\ve{x}$ and $G(\ve{z},\tilde{\ve{x}})$. 

\section{Generalized Speech Enhancement}
\label{sec:get}


In this work, we depart from the classical denoising task and consider a more general class of enhancement problems. More specifically, we consider the problem of reconstructing speech that has been degraded by a number of (different) signal manipulations, each of them being potentially highly perceptually harmful. We propose to mix up such manipulations together with several speaker identities at training time, with the objective of being able to train generative algorithms to recover multiple (and diverse) speech components simultaneously. 
As a first approximation towards this generalized speech enhancement task, we consider the following signal manipulations:
\begin{itemize}
    
\item \textbf{Whispered speech:} This is the distortion is similar to the one introduced in the WSEGAN work~\cite{pascual2018whispered}. However, it differs in the fact that 
instead of using magnetic sensors, 
we synthesize whispering speech by encoding the clean speech with a vocoder, removing the log-F0 (i.e. making all frames unvoiced), and recovering the signal into a version that whispers, hence artificially removing voicing. The vocoder used is Ahocoder with the default parameters~\cite{erro2011improved}. 

\item \textbf{Bandwith reduction:} We downsample the audio signals with different severity factors, ranging from $\times2$ to $\times8$, thus reducing the bandwidth. Then the generalized enhancement model must reconstruct entire frequency bands, according to the clean signals seen in training.

\item \textbf{Chunk removal:} For the parts of the waveform that contain speech, a random number of chunks are subtracted by inserting silences. Thus, we cancel the signal at random temporal sections that contain speech. The length of a silence is sampled from one of two distributions $\mathcal{N}(0.05,0.025)$, and $\mathcal{N}(0.1,0.05)$ (numerical values in seconds).

\item \textbf{Clipping:} The waveform is clipped globally by different severity factors relative to the maximum absolute peak of the whole utterance (e.g., 30\%). The regenerated signal thus has to re-condition the signal inside a proper, non-distorted range of amplitudes.

\end{itemize}

\section{Acoustic Mapping Discriminator}

To better deal with the added difficulty that the previous signal manipulations can introduce to the speech signal, we propose two modifications to the existing SEGAN pipeline: the addition of two acoustic losses and the introduction of a two-stage training schedule for applying such losses. 

\subsection{Acoustic Losses}
\label{sec:aco_losses}

In general, $D$ can be understood as a learnable loss function, where the realistic features we want to generate are implicit in the back-propagation, and $G$ gets better because of $D$'s gradient flows~\cite{goodfellow2014generative}. This builds a need for $D$ to be a competitive feature extractor, so that the better the features extracted by $D$, the better $G$ can capture reality. 
The use of auxiliary classifying labels in $D$ is generally helpful (Sec.~\ref{sec:intro}). Additionally, multi-task setups can boost generative modeling performance, where multiple factors of the signal are predicted at the output of the generative model to enforce modeling better perceptual qualities of the signal~\cite{oord2017parallel}. This is applicable to the SEGAN setup where the discriminator only learns features that lead to a \texttt{real} or \texttt{fake} decision, but is not concerned about specific factors of the signal that are important to remark the reconstructed speaker identity, prosody, or contents. Note that we can also have some additional regularization loss in the output of $G$ too, aggregated to the adversarial loss coming from $D$. For instance the first version of SEGAN used an $L_1$ regularization term that made $G$ output a zero-centered signal~\cite{pascual2017segan}, and the posterior WSEGAN implementation substituted it by a less restrictive power loss~\cite{oord2017parallel, ping2018clarinet}, enforcing proper energy allocation in the frequency bands.

Following the aforementioned multi-task improvement criteria, we propose the use of an additional acoustic loss that opens a new branch in the end of $D$. This makes $D$ necessarily aware of additional acoustic components that can inform $G$ about important mistakes when mismatching identities, intonations, or contents in the reconstruction without necessarily injecting any additional conditioning in the input of $G$ besides the distorted signal $\tilde{\ve{x}}$. This is advantageous in terms of avoiding restricting the generalization of $G$, as we do not inject any code limited by a set of possible classes to specify identities, distortion types, or any other kind of information. This new $D$ output branch grows from a certain level in the convolutional structure, where the time decimation factor is 256, thus each time-step corresponds to 16\,ms stride over the waveform (emulating a short-time Fourier transform sliding window). At each output frame, it predicts a concatenation of log-power spectral (LPS) bins, Mel-frequency cepstral coefficients (MFCC), and prosodic features. We design it using this collection of features as we hypothesize they properly convey different levels of acoustic cues, ranging from power allocation expressing identity, to content and intonation clues. This branch is just in addition to the binary output neuron that tells \texttt{real} and \texttt{fake}, and is only active when this binary activation is \texttt{real} with input $\ve{x}$ or $\tilde{ve{x}}$. The proposed acoustic loss in $D$ and power regularization in $G$ redefines the loss function as
\begin{equation*}
\begin{split}
  \underset{G}\min~V(G) & = \frac{1}{2}\E_{\ve{z}\sim p_{\ve{z}}(\ve{z}),\tilde{\ve{x}}\sim p_{\text{data}}(\tilde{\ve{x}})}[(D(G(\ve{z},\tilde{\ve{x}}),\tilde{\ve{x}}) - c)^{2}] \\
  & + \frac{1}{2}\E_{\ve{z}\sim p_{\ve{z}}(\ve{z}),\ve{x},\tilde{\ve{x}}\sim p_{\text{data}}(\ve{x}, \tilde{\ve{x}})}[(\delta(G(\ve{z},\tilde{\ve{x}}),\tilde{\ve{x}}) - \Theta({\ve{x}}))^{2}]\\
  & + \alpha \E_{\ve{z}\sim p_{\ve{z}}(\ve{z}),\ve{x},\tilde{\ve{x}}\sim p_{\text{data}}(\ve{x}, \tilde{\ve{x}})}[\Phi(G(\ve{z},\tilde{\ve{w}})) - \Phi(\ve{x})] ,\\
  \underset{D}\max~V(D) & = \frac{1}{4}\E_{\ve{x},\tilde{\ve{x}}\sim p_{\text{data}}(\ve{x}, \tilde{\ve{x}})}[(D(\ve{x},\tilde{\ve{x}}) - b)^{2}] \\
    & + \frac{1}{4}\E_{\ve{x},\tilde{\ve{x}}\sim p_{\text{data}}(\ve{x}, \tilde{\ve{x}})}[(\delta(\ve{x},\tilde{\ve{x}}) - \Theta({\ve{x}}))^{2}] \\
    & + \frac{1}{4}\E_{\ve{z}\sim p_{\ve{z}}(\ve{z}),\tilde{\ve{x}}\sim p_{\text{data}}(\tilde{\ve{x}})}[(D(G(\ve{z},\tilde{\ve{x}}),\tilde{\ve{x}}) - a)^{2}]\\
    & + \frac{1}{4}\E_{\ve{x},\ve{x}^{r}\sim p_{\text{data}}(\ve{x})}[(D(\ve{x},\ve{x}^r) - a)^{2}],
\end{split}
\label{eq:lsgan_aco}
\end{equation*}
where $\ve{x}$, $\tilde{\ve{x}}$, $\ve{x}^r$, $\ve{z}$, $a$, $b$, and $c$ are the same as in Eq.~\ref{eq:lsgan}, $\Phi(\ve{x})$ corresponds to the short-time Fourier transform magnitude in dBs (20\,ms windows, 10\,ms stride, and 2048~bias), and $\alpha=10^{-3}$ is a weighting term~\cite{pascual2018whispered}. The $\delta$ function is the discriminator output at some intermediate layer of our choice, and $\Theta(\ve{x})$ is the feature extractor of LPS, MFCC and prosodic features (see sec.~\ref{sec:segan_setup}). Hereafter, we will refer to the joint power loss regularization and acoustic loss as acoustic losses.

\begin{figure}
    \centering
    \includegraphics[width=0.9\linewidth]{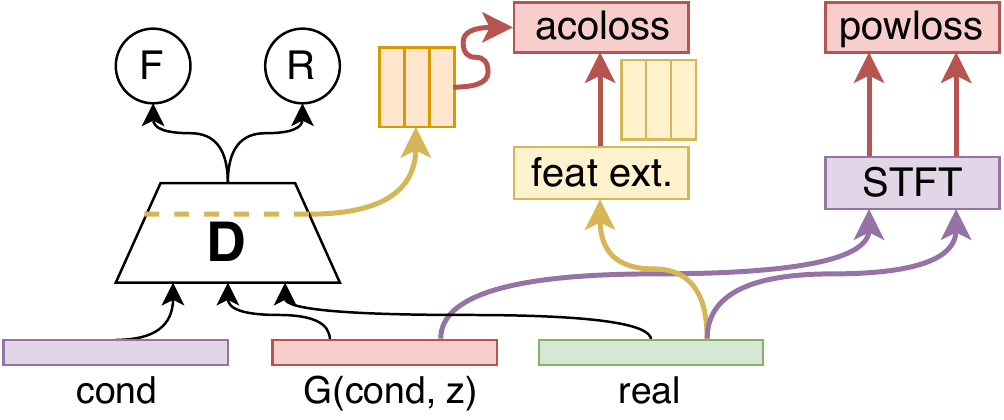}
    \caption{Setup of new multi-task framework with acoustic loss, power loss, and adversarial outcomes (F and R).}
    \label{fig:acoloss}
\end{figure}

\subsection{Adversarial Pre-Training}

Whereas SEGAN with only an adversarial loss is stable and learns in a steady equilibrium (eq.~\ref{eq:lsgan}), the addition of the acoustic losses induces a particular unbalance effect during training. The addition of these terms makes $D$ learn quicker and converge faster in both losses, conversely making $G$ converge slower. Importantly, both $G$ and $D$ should maintain an equilibrium learning from each other, so making one of them quickly better discourages the other to perform properly~\cite{goodfellow2014generative}. 
We hypothesize that scheduling the learning of the discriminator as a two-stage process makes the addition of these acoustic losses more effective for it first gets high-level representations classifying, and then focuses on specific speech properties by doing regressions. 
Hence we first do an adversarial warm up in $D$ with Eq.~\ref{eq:lsgan} formulation, and then add the acoustic losses.

\section{Experimental Setup}

\subsection{Dataset}
\label{sec:dataset}

We employ the VCTK Corpus for our experiments~\cite{veaux2016superseded}. We select 80 speakers for training and 14 for test. We trim large silence regions to 100\,ms with the help of a voice activity detector. After this, we get roughly 20\,h of training speech and 3\,h of test speech. We incorporate the aforementioned distortions (Sec.~\ref{sec:get}) with an online process jointly working with training. When we construct a training minibatch, we (1) load the clean utterance, (2) get a random $\approx$1\,s chunk (16384 samples) inside the utterance, and (3) apply a series of transforms that get activated independently under a probability $p=0.4$. This way one or two distortions coincide often, and zero (thus purely autoencoder) or more than two coincide less. Table~\ref{tab:histogram_tranforms} shows this activation distribution with up to the 4 mentioned combinations. In addition to being active or not, each transform has a certain severity level or factor, as mentioned before (sec.~\ref{sec:get}). The only transform with just one level of severity is the whispering transform. Hence every time a transformation is activated in the pipeline, one of the possible factors is randomly selected. These possibilities are: clip factors of 30\%, 40\% and 50\%, signal resampling factors of 2, 4, and 8 and, in the case of chunk removal, we allow the system to zero out up to $5$ chunks from speech regions. Then, inside each region, the length of the chunk is sampled from one of the two mentioned Gaussian distributions (sec.~\ref{sec:get}).

\begin{table}[t!]
\centering
\caption{Frequency of number of training active transformations using $p=0.4$.}
\label{tab:histogram_tranforms}
\begin{tabular}{p{2.5cm}|c|c|c|c|c}
    \hline 
     Num. of transforms & 0 & 1 & 2 & 3 & 4 \\
     \hline
     Relative frequency & 0.14 & 0.34 & 0.33 & 0.15 & 0.04 \\
     \hline
\end{tabular}
\end{table}

For the test split, we have two different setups, one for an objective evaluation and another one for a subjective one. For the objective test set, we proceed as with the training set. For the subjective test, we generate a special split with two subsets in order to account for two different effects present in the system outcomes. The first subset is designed to focus on generated speaker identity (\textit{SpkID} pool). To do so, we just use the most severe version of bandwidth reduction ($\times$8) and the whispering manipulations, so the ones that affect the most this feature. Then, for each test speaker we degrade four utterances, two with bandwidth reduction and two with whispering. The second subset focuses on the generated speech naturalness (\textit{Nat} pool). To evaluate this aspect, we pick 4~test speakers (two male and two female) and apply clipping, resampling, and whispering to 6 utterances in total (two utterances per distortion). Both clipping and resampling have severity factors 30\% and $\times$8, respectively. 

\subsection{Experiments}

We experiment with three models applied over the aforementioned data. Firstly, we use SEGAN in a plain adversarial setup with the LSGAN loss as in Eq.~\ref{eq:lsgan}, where the only signal comes from $\texttt{real}$ or $\texttt{fake}$ decisions. This is the baseline, which is trained for 400~epochs with two-timescale update rule (TTUR) learning rates~\cite{heusel2017gans} $\eta_D = 4\cdot 10^{-4}$ and $\eta_G = 1\cdot 10^{-4}$. Secondly, we apply the acoustic losses presented in Sec.~\ref{sec:aco_losses}. This system trains also for 400~epochs, but the learning rates are balanced equal $\eta_D = \eta_G = 5\cdot 10^{-5}$, because the discriminator already has an advantage with the extra signals and TTUR involved a noisier result on $G$. We name this approach SEGAN-Aco. Finally, we include our improved proposal, for which we pre-train the system over 100~epochs the same way we do with the baseline, and then we activate the acoustic losses for the remaining 300~epochs, also lowering the learning rates to $\eta_D = \eta_G = 5\cdot 10^{-5}$. We name this approach SEGAN-PTAco.

\subsection{Evaluation}

To assess each model result we perform two evaluations. First, we run a number of objective distortion metrics, often applied in speech synthesis problems: Mel cepstral distortion (MCD; in dB)~\cite{kubichek1993mel}, F0 root mean squared error (RMSE; in Hz), and the voiced/unvoiced frame prediction error (UV)~\cite{pascual2016multi}. These errors give us a first clue on how close is each system to the clean original signal in terms of content, identity, and intonation.

Given the importance of perceptual scores, we also conduct a subjective evaluation with 26~listeners. This has two stages, with two tasks to be evaluated by the listeners: speaker identification and naturalness rating. For speaker identification, listeners are asked to determine how close a reconstructed signal is towards the original speaker identity. They are presented with 4~randomly-selected utterances out of the \textit{SpkID} pool (Sec.~\ref{sec:dataset}). For each utterance, the clean reference is shown, as well as the four systems to be rated: (1) the degraded input signal to $G$, (2) the SEGAN baseline, (3) SEGAN-Aco, and (4) SEGAN-PTAco. The rating is as simple as ordering them by preference, such that position 1 is for closest system towards the reference and position 4 is the furthest one. For naturalness rating, 6~utterances taken randomly out of the \textit{Nat} pool (Sec.~\ref{sec:dataset}) are shown to each listener. For each utterance, the four different aforementioned systems are shown and asked to be ranked from most natural (1) to least natural (4). There is no reference shown for this case as there is no similarity trait like a speaker identity, it is only comparing the perceptual quality of the synthesized utterances. In all subjective evaluations, systems are shown in random order at every utterance.

\subsection{SEGAN Setup}
\label{sec:segan_setup}
We use the same kernel widths, strides, and feature map configurations as in the WSEGAN work~\cite{pascual2018whispered}. These are kernels of width 31 in all convolutional layers for both $G$ and $D$. The feature maps are incremental in the encoder and decremental in the decoder, having $\{64, 128, 256, 512, 1024, 512, 256, 128, 64, 1\}$ in $G$ and $\{64, 128, 256, 512, 1024\}$ in $D$. The discriminator is then the one diverging from its predecessors in earlier works~\cite{pascual2017segan, pascual2018whispered}. Firstly, it has a multi-layer perceptron (MLP) with 16384 inputs (1024$\times$16 feature map, unrolled), 256~hidden PReLU units, and the single output for \texttt{real} and \texttt{fake} predictions. Secondly, we find the acoustic branch for SEGAN-Aco and SEGAN-PTAco models at the fourth convolutional layer, which outputs 512$\times$64 feature maps for an input of approximately 1\,s at 16\,kHz sampling rate. These corresponds to 64~frames, each one injected into the acoustic prediction MLP of 128~hidden PReLU units and 277~linear outputs. These outputs predict 257~log-power spectral bins, 16~MFCCs, 1~log-F0 value of the frame, 1~voiced/unvoiced frame flag, 1~frame energy coefficient, and 1~frame zero crossing rate. Mini-batches of 150 samples are used for all models. Additionally, $D$ implements spectral normalization to avoid sudden exploding gradients leading to training collapse~\cite{zhang2018self, miyato2018spectral} and phase shift of $5$ to reduce high-frequency artifacts in the output of $G$~\cite{donahue2018synthesizing}. All the experiments were developed with the public SEGAN framework at \texttt{https://github.com/santi-pdp/segan\_pytorch}. 

\section{Results}

Objective results are shown in Table~\ref{tab:obj_metrics}. We can see that the distorted signals themselves have the noisiest behaviors, with large variances in all error metrics. This is expectable, as we have a wide range of distortion conditions. We can see how, in all metrics, SEGAN-PTAco is the best system and the least noisy one. A lower MCD can be assumed to correlate with better content preservation, identity reconstruction, and naturalness generation~\cite{pascual2016deep}. The lower values obtained for F0-RMSE and UV error also typically denote better intonation schemes matching the input identity with prosodic contents.

\begin{table}[t]
\centering
\caption{Objective metrics for the different systems (standard deviation into parenthesis). For each metric, lower is better.}
\label{tab:obj_metrics}
\setlength{\tabcolsep}{6pt}
\begin{tabular}{l|c|c|c}
    \hline 
    Model & MCD~[dB] & RMSE~[Hz] & UV~[\%] \\
     \hline
     Distorted & 7.6~(5.6) & 52.9~(75.9) & 22.3~(25.5) \\
     SEGAN & 8.2~(3.5) & 37.2~(43.5) & ~\,9.1~(11.4) \\
     SEGAN-Aco & 7.4~(2.7) & 37.5~(53.5) & 20.0~(23.4) \\
     SEGAN-PTAco & \textbf{6.5~(2.7)} & \textbf{22.7~(28.5)} & \textbf{~\,5.8~(~~6.3)} \\
     \hline
\end{tabular}
\end{table}

Subjective results are shown in Table~\ref{tab:obj_metrics}. In general, they also highlight the importance of both the acoustic losses and the two-stage adversarial training schedule. We can see that the baseline and SEGAN-Aco are clustered together with the distorted signals for the Speaker ID test. This implies that, for both bandwidth extension and dewhispering problems, neither systems reconstruct a proper speaker identity from the input signal. A qualitative listening in this case tells us that the baseline imposes an arbitrary identity to the reconstruction, and SEGAN-Aco sounds robotic and muffled. In contrast, SEGAN-PTAco is consistently ranked above the distorted signal, and we can clearly hear a competitive identity consistency with the conditioning. In the naturalness task, the baseline and SEGAN-Aco systems do better than the distorted signals. This was expected, as they involve an enhancement process that removes very noticeable artifacts and recover speech nuances. Nevertheless, SEGAN-PTAco is still the best system by a large margin. Some audio samples are available at \url{http://veu.talp.cat/gsegan/}.

\begin{table}[t]
\centering
\caption{Subjective mean ranking score (lower is better).}
\label{tab:obj_metrics}
\setlength{\tabcolsep}{10pt}
\begin{tabular}{l|c|c}
    \hline
    Model & Speaker ID & Naturalness \\
     \hline
     Distorted & 2.87~(1.07) & 3.00~(1.04) \\
     SEGAN & 2.85~(1.03) & 2.69~(1.04) \\
     SEGAN-Aco & 2.88~(0.83) & 2.78~(0.92) \\
     SEGAN-PTAco & \textbf{1.41~(0.74)} & \textbf{1.52~(0.84)} \\
     \hline
\end{tabular}
\end{table}

\section{Conclusion}

In this work, we propose the application of speech enhancement GANs in a more generalized signal recovery framework. We first introduce four aggressive distortions applied in our experiments emulating aggressive distortions to the speech signals. Next, we introduce a new acoustic regression component in the SEGAN $D$ and, in addition, we propose a two-stage adversarial training method to make the new acoustic regression component reach an appropriate performance. Objective and subjective results are provided, showing how both the regression losses and and the two-stage schedule outperform the regression losses alone and the purely adversarial approach.

\section{Acknowledgements}
This research was partially supported by the project TEC2015-69266-P (MINECO/FEDER, UE). We deeply thank the participants of the subjective evaluation.

\newpage
\bibliographystyle{IEEEtran}
\bibliography{mybib}


\end{document}